\documentclass[12pt,a4paper]{article}
\usepackage{amsmath}
\usepackage{amsxtra}
    \usepackage{amstext}
    \usepackage{amssymb}
    \usepackage{latexsym}
    \usepackage{graphicx}
\usepackage{color}
\usepackage{graphics}

\newcommand{\hepth}[1]{{\tt hep-th/#1}}

\newcommand{\nn}{\nonumber}

\newcommand{\p}{\vspace{6pt}\noindent}

\advance\textheight by \topskip
\makeatletter

\@addtoreset{equation}{section}
\def\section{\@startsection {section}{1}{\z@}{-8.5ex plus -1ex minus
 -.2ex}{3.3ex plus .2ex}{\large\bf}}
\def\subsection{\@startsection{subsection}{2}{\z@}{-3.25ex plus
 -1ex minus -.2ex}{1.5ex plus .2ex}{\bf}}
\def\subsubsection{\@startsection{subsubsection}{3}{\z@}{-3.25ex plus%
 -1ex minus -.2ex}{1.5ex plus .2ex}{\sl}}

\begin{document}

\begin{titlepage}
\vspace*{-2cm}
\begin{flushright}
\end{flushright}

\vspace{0.3cm}

\begin{center}
{\Large {\bf }} \vspace{1cm} {\Large {\bf A transmission matrix for a fused pair of integrable defects in the sine-Gordon model}}\\
\vspace{1cm} {\large  E.\ Corrigan\footnote{\noindent E-mail: {\tt
edward.corrigan@durham.ac.uk}} and
C.\ Zambon\footnote{\noindent E-mail: {\tt cristina.zambon@durham.ac.uk}} \\
\vspace{0.3cm}
{\em Department of Mathematical Sciences \\ University of Durham, Durham DH1 3LE, U.K.}} \\

\vspace{2cm} {\bf{ABSTRACT}}
\end{center}

\p Within the quantum sine-Gordon model a transmission matrix describing the scattering of a soliton with a fused pair of integrable defects is proposed. The result is consistent with the classical picture of scattering and highlights the differences between two defects located at separated points and two defects fused at the same point. Moreover, the analysis reveals how, for certain choices of parameters, both the soliton-soliton and the lightest-breather-soliton S-matrices of the sine-Gordon model are embedded within the transmission matrix, supporting an interpretation in which  defects may be regarded as soliton constituents.

\vfill
\end{titlepage}

\section{Introduction}

Gradually, it has been realised that defects play a role even in integrable field theory. It was pointed out by Delfino, Mussardo and Simonetti \cite{Delf94} that impurities might be a fruitful area to study despite the fact they were likely to be strongly constrained. Though one might expect a typical impurity to be both transmitting and reflecting, compatibility with the bulk S-matrix in an integrable quantum field theory with non-trivial bulk scattering would require the impurity to be purely transmitting. An example of a suitable transmission matrix was provided in \cite{Konik97}. Some time later, Lagrangian-based examples of integrable defects were introduced in \cite{bcz2003} where a typical impurity is a field discontinuity together with a set of sewing conditions connecting the fields to either side of the discontinuity. Further work on these defects can be found in \cite{bcz2005,others,Bajnok} while a framework within which defects allow both reflection and transmission can be found in \cite{Caudrelier:2004hj}; for additional issues concerning topological defects, see \cite{Bajnok:2009hp}.

\p
Recently, it has been pointed out in \cite{cz2009} how it is possible to generalise the classical description of integrable defects previously investigated in \cite{bcz2003} (hereafter called type I) to encompass defects within the Tzitz\'eica or $a_2^{(2)}$  Toda model.\footnote{\ Often referred to as the Bullough-Dodd or Mikhailov-Zhiber-Shabat equation.} To achieve this an additional degree of freedom was added at the location of the defect (and in future these will be called type II). Within the sine-Gordon model this mechanism can be interpreted as being the result of `fusing' two defects of the kind elaborated in \cite{bcz2003, bcz2005}. At least, this is the picture suggested by studying the classical scattering of a soliton with a type II defect. In this article, the aim is to explore the extent to which this interpretation can be seen to be valid within the quantum sine-Gordon theory. It was remarked in \cite{bcz2005} that two slightly separated defects in combination behave classically like a soliton when scattering with a genuine soliton. This observation provided further motivation to explore the manner in which this behaviour might appear within the quantum field theory and to discover the precise terms on which a type II defect might resemble a soliton.

\p
As mentioned above, some years ago Konik and LeClair found a solution of the `triangle' equations representing the compatibility relations between the transmission matrix and the bulk S-matrix \cite{Konik97}. This was rederived and analysed from a semi-classical point of view in \cite{bcz2005}, and there it was concluded, on the basis of the evidence gathered, that it was the appropriate solution for the scattering of a soliton with the basic type I defect of the sine-Gordon theory. However, there are other infinite dimensional solutions to the triangle equations, some of them obtained straightforwardly by taking products of Konik-LeClair solutions, and others that are more subtle in a manner that will be explained below. It will be argued it is the latter that most appropriately describe the scattering of  solitons with a type II defect.

\p
Interestingly, the bulk scattering matrix reappears as a special case within the set of possible transmission factors. For the special choice of defect parameters for which this occurs it could be argued that a projection of the generalised defect behaves exactly like a soliton. Put another way: this fact supplies a little more evidence that the generalised defect supplies an alternative description of a soliton in terms of `sewing conditions' rather than the more usual description using an extended, smooth, field configuration solving the classical equations of motion.

\p
The plan of the article is as follows. In section 2 some of the basic facts concerning defects will be described and in section 3 a new solution to the triangle equations will be presented. Properties of the latter will be provided in section 4, together with its connections with the S-matrix. Finally, section 5 contains some concluding remarks and future directions.

\section{The classical setting}
\label{classicalsetting}

\p The starting point is the Lagrangian density for a a type II defect:
\begin{equation}\label{defectlagrangian}
{\cal L}=\theta(-x){\cal L}_u + \theta(x){\cal L}_v +\delta(x)\left(2 q\lambda_t -{\cal D}(\lambda,u,v)\right),
\end{equation}
where $u,v$ are fields in $x<0$ and $x>0$, respectively, separated by a defect at $x=0$, and $q=(u(0,t)-v(0,t))/2$ represents the discontinuity at $x=0$. The quantity $\lambda(t)$ is confined to $x=0$ and the consequences of the defect, in the sense of sewing conditions for the fields $u$ and $v$ at $x=0$ are encoded by the defect potential ${\cal D}$. The fields $u,v$ are described within their domains by the sine-Gordon Lagrangian densities ${\cal L}_u$, ${\cal L}_v$ with potentials
$$U(u)=2-(e^u+e^{-u}),\qquad V(v)=2-(e^v+e^{-v}),$$
respectively. In this section, the coupling constant and mass scale have been removed to simplify the formulae but may be reinserted by rescaling coordinates and fields, if so desired. Note that the fields $u$ and $v$ may be generally complex but are  purely imaginary for the sine-Gordon model.
Finally, the defect potential is:
\begin{eqnarray}\label{defectpotential}
{\cal D}(\lambda,p,q)&=&-\sqrt{2}\sigma (e^{p/2+\lambda/2}(e^{q/2-\tau}+e^{-q/2+\tau})+ e^{-p/2-\lambda/2}(e^{q/2+\tau}+e^{-q/2-\tau}))\nonumber \\
&&-\frac{\sqrt{2}}{\sigma}(e^{p/2-\lambda/2}(e^{q/2+\tau}+e^{-q/2-\tau})
+e^{-p/2+\lambda/2}(e^{q/2-\tau}+e^{-q/2+\tau})),\nonumber \\
&=&f(p+\lambda, q)+g(p-\lambda, q),\qquad p=(u+v)/2,
\end{eqnarray}
where $\sigma$ and $\tau$ are two free parameters.

\p Besides the bulk equations for the two fields in their respective domains, eq\eqref{defectlagrangian} provides a set of defect conditions linking the fields at $x=0$. In terms of $\lambda$, $p$ and $q$ they are:
\begin{equation}\label{defectconditions}
2 q_x=-\mathcal{D}_p,\qquad 2p_x-2\lambda_t=-\mathcal{D}_q,\qquad 2q_t=-\mathcal{D}_\lambda.
\end{equation}
In eqs\eqref{defectconditions}, the quantities $q_x$ and $p_x$ are constructed using limiting values of $u_x$ and $v_x$ as $x\rightarrow 0$ from below or above zero, respectively. In \cite{cz2009} it was shown that for the system described by \eqref{defectlagrangian} both energy and momentum are conserved. Moreover, evidence to support the conjecture that the system is integrable was presented. However, it should be remarked that although the defect Lagrangian density used in \cite{cz2009} appears to differ from  \eqref{defectlagrangian}, they are effectively the same. Indeed, that the two formulations describe the same problem can be seen by performing the following shift in $\lambda$,
\begin{equation}\label{Lambdashift}
\lambda\rightarrow -\lambda/2-\ln(e^{q/2+\tau}+e^{-q/2-\tau})+p/2+\ln 2/2,
\end{equation}
as a result of which the defect Lagrangian density used as the starting point in \cite{cz2009} is transformed into \eqref{defectlagrangian} with defect potential \eqref{defectpotential}. This redefinition of the field $\lambda$ can be used to compare formulae obtained via the two different formulations.
The choice \eqref{defectlagrangian} for the defect Lagrangian density is more suitable for its generalization to other relativistic field theories such as the Toda models. In addition, it is straightforward to see, as was noted  in the case of the sine-Gordon model \cite{cz2009}, that the type II defect can be interpreted as two type I defects fused at the same point on the $x$-axis. In fact, adding together two type I defect terms, one for a pair of fields $u$, $\lambda$, and the other for a pair of fields $\lambda$, $v$ gives
$$\delta(x)\left(\frac{u \lambda_t-\lambda u_t}{2}+\frac{\lambda v_t-v \lambda_t}{2}-\mathcal{D}_1(u,\lambda)-\mathcal{D}_2(\lambda,v)\right),$$
with
\begin{eqnarray}
\mathcal{D}_1(u,\lambda)&=&-\sqrt{2}\,\sigma_1(e^{u/2+\lambda/2}+e^{-u/2-\lambda/2})
-\frac{\sqrt{2}}{\sigma_1}(e^{u/2-\lambda/2}+e^{-u/2+\lambda/2}),\nonumber\\
\mathcal{D}_2(\lambda,v)&=&-\sqrt{2}\,\sigma_2(e^{\lambda/2+v/2}+e^{-\lambda/2-v/2})
-\frac{\sqrt{2}}{\sigma_2}(e^{\lambda/2-v/2}+e^{-\lambda/2+v/2}).\nonumber
\end{eqnarray}
The type II defect term in \eqref{defectlagrangian} is obtained on identifying parameters as follows:
$$\sqrt{\sigma_1\sigma_2}=\sigma\equiv e^{-\eta},\qquad \sqrt{\frac{\sigma_1}{\sigma_2}}=e^{-\tau}.$$

\p However, the complete Lagrangian density \eqref{defectlagrangian} is not equivalent to the sum of the Lagrangian densities for two type I defects located at the same point on the $x$-axis because the bulk Lagrangian density for the field $\lambda$ has been suppressed. This suppression has consequences and to emphasize the distinction between the type I and type II defects for the sine-Gordon model it should be noted that the type II defect cannot be split into two separated type I defects. The reason for this is clear: for the separated defects four sewing conditions would be needed - two from each type I defect - yet the Lagrangian density \eqref{defectlagrangian} only provides three conditions. This subtle difference between a type II and two type I defects enters also in the quantum context as will be explained in the next sections. At present, it suffices to notice that once two type I defects are fused a soliton cannot propagate between them, and therefore the transfer of topological charge between them is suppressed. Another  fact concerns the  time-independent field configurations. Since the sine-Gordon energy density is bounded below, the ground states of zero total energy can be chosen to be multiples of $2\pi$. Field configurations corresponding to odd multiples of $\pi$, even within a finite domain have a finite energy proportional to the size of the domain. Only when these field configurations are confined to a single point do they become degenerate with the ground states.\footnote{\ In the context of the type II defect for a single massive scalar field, a bound state with energy degenerate with the ground state has been noted \cite{cz2009}} Finally, it is worth recalling that, as shown for the Tzitz\'eica equation in \cite{cz2009}, the type II defect can occur in models that cannot support  type I defects. For these reasons, the two types of defect will be regarded as distinct.

\p As a prelude to the quantum analysis, it is useful to understand the classical behaviour of a soliton  traversing a defect and for this purpose the defect conditions \eqref{defectconditions} will be used. Generally, at the defect, the values of the fields $u$ and $v$ do not match and the difference $u(0,t)-v(0,t)$ represents the defect discontinuity. At early or late times this can be either $0$ or $2\pi i$ modulo $4\pi i$. These two possibilities are different and  will be treated separately. First, consider a defect with no discontinuity. The asymptotic behaviour of the solitons $u(0,t)$ and $v(0,t)$ as $t\rightarrow -\infty$ can be chosen, for example, to be  $u=v=2\pi i$. Moreover, the defect conditions \eqref{defectconditions} must be satisfied and they force the initial value of $\lambda$ to be either $0$ or $2\pi i$, also modulo $4\pi i$. The general form of the soliton solutions for the fields $u$ and $v$ is known and conveniently summarised implicitly by:
\begin{equation}\label{setting1}
e^{u/2}=\frac{1+  E}{1-  E},\quad E=e^{ ax+bt+c}, \quad a=\sqrt{2}\cosh\theta,\quad b=-\sqrt{2}\sinh\theta,
\quad e^{v/2}=\frac{1+z E}{1-z E},
\end{equation}
where $e^c$ is a purely imaginary constant and $z$ represents the delay of the outgoing soliton. Note that these expressions for $u$ and $v$ agree with their chosen initial values at $x=0$. Then, the defect conditions \eqref{defectconditions} may be used to find the delay $z$ and the general form for the field $\lambda(t)$. Since, as noticed previously,  $\lambda$ may have one of two different initial values,  two different delays are expected together with two expressions for the field $\lambda(t)$ as a function of time, and details have been provided in ref\cite{cz2009}. Here,  the delays will be summarised, together with the total energies associated with  particular configurations.  If $\lambda (-\infty)=0$:
\begin{eqnarray} z_1&=&\tanh\left(\frac{\eta+\tau-\theta}{2}\right)\,\tanh\left(\frac{\eta-\tau-\theta}{2}\right), \nonumber\\
\mathcal{E}_1&=&8\sqrt{2}\cosh\theta+[8\sqrt{2}\cosh{\eta}\cosh{\tau}].
\label{4piconfiguration2}
\end{eqnarray}
where the part within square brackets is the defect contribution. On the other hand, if $\lambda (-\infty)=2\pi i$:
\begin{eqnarray} z_2&=&\coth\left(\frac{\eta+\tau-\theta}{2}\right)\,\coth\left(\frac{\eta-\tau-\theta}{2}\right), \nonumber\\
\mathcal{E}_2&=&8\sqrt{2}\cosh\theta+[- 8\sqrt{2}\cosh{\eta}\cosh{\tau}], \label{4piconfiguration1}
\end{eqnarray}
A similar analysis can be performed when the defect has a $2\pi i$ discontinuity. This can be realized by choosing  the initial $u$ and $v$ at $x=0$ to be $2\pi i$ and $0$, respectively. The soliton expressions for the two fields compatible with these initial conditions are:
\begin{equation}\label{setting2}
e^{u/2}=\frac{1+  E}{1-  E},\quad E=e^{ ax+bt+c}, \quad a=\sqrt{2}\cosh\theta,\quad b=-\sqrt{2}\sinh\theta,
\quad e^{v/2}=-\frac{1+z E}{1-z E}.
\end{equation}
The defect conditions again determine two possible initial values for the field $\lambda$.
If $\lambda (-\infty)=0$:
\begin{eqnarray} z_3&=&\coth\left(\frac{\eta+\tau-\theta}{2}\right)\,\tanh\left(\frac{\eta-\tau-\theta}{2}\right), \nonumber\\
\mathcal{E}_1&=&8\sqrt{2}\cosh\theta+ [8\sqrt{2}\sinh{\eta}\sinh{\tau}], \label{2piconfiguration1}
\end{eqnarray}
but, if $\lambda (-\infty)=2\pi i$:
\begin{eqnarray} z_4&=&\tanh\left(\frac{\eta+\tau-\theta}{2}\right)\,\coth\left(\frac{\eta-\tau-\theta}{2}\right), \nonumber\\
\mathcal{E}_2&=&8\sqrt{2}\cosh\theta+[-8\sqrt{2}\sinh{\eta}\sinh{\tau}].
\label{2piconfiguration2}
\end{eqnarray}
Note, as pointed out in \cite{cz2009}, the delays may be negative,  zero, or diverge, according to the relative values of the soliton rapidity and the defect parameters. These represent an opportunity for the soliton to convert to an antisoliton in the first case or to be absorbed by the defect in the other two cases.
If the defect parameters are taken to be real, the lowest energy configuration will be realized in the presence of a defect with a $0$ modulo $4\pi i$ discontinuity. In addition, assuming the defect parameters are positive, a soliton crossing such a defect will acquire a  delay $z_2$. Such a configuration is  stable and in the quantum context will be described by a unitary transmission matrix.

\p It is useful to calculate the transmission factor for the sine-Gordon defect problem linearized about this particular, stable, configuration. The result is:
\begin{equation}\label{TFlinearizedproblem}
T_2=-\frac{\sinh\left(\frac{\theta-\eta+\tau}{2}-\frac{i\pi}{4}\right)\sinh
\left(\frac{\theta-\eta-\tau}{2}-\frac{i\pi}{4}\right)}
{\sinh\left(\frac{\theta-\eta+\tau}{2}+\frac{i\pi}{4}\right)\sinh
\left(\frac{\theta-\eta-\tau}{2}+\frac{i\pi}{4}\right)}.
\end{equation}
This will coincide with the classical limit of the transmission factor for the lightest breather calculated in the next section.

\p Finally, it is interesting to discuss the possibility for interpreting the type II defect as a soliton. For that, the setting \eqref{setting2} must be chosen since there the defect appears with the appropriate $2\pi i$ discontinuity. Consider, for instance, the solution \eqref{2piconfiguration1}. The defect parameters are free constants that can be chosen to be complex. Set $\eta+\tau=\vartheta\pm \pi i$ and $\eta-\tau=\vartheta$. Then, the defect contribution to the energy becomes equal to the energy of a soliton with rapidity $\vartheta$, and the delay $z_3$ becomes $\tanh^2(\vartheta-\theta)/2$,  which is the classical delay factor for two scattering solitons with rapidities $\vartheta$ and $\theta$. In addition, the transmission factor of the sine-Gordon defect problem linearized about this configuration reads
\begin{equation}
T_3=\frac{\sinh\left(\frac{\theta-\eta+\tau}{2}-\frac{i\pi}{4}\right)
\sinh\left(\frac{\theta-\eta-\tau}{2}+\frac{i\pi}{4}\right)}
{\sinh\left(\frac{\theta-\eta+\tau}{2}+\frac{i\pi}{4}\right)
\sinh\left(\frac{\theta-\eta-\tau}{2}-\frac{i\pi}{4}\right)},
\end{equation}
which, with the resetting of the defect parameters suggested above becomes
\begin{equation}\label{classicalsolitonbreather}
T_3=\frac{\sinh(\theta-\vartheta)+i}
{\sinh(\theta-\vartheta)-i}.
\end{equation}
This coincides with the classical limit of the transmission factor of a soliton scattering with the lightest breather, as will be seen in the next sections.
Similar considerations hold for the solution \eqref{2piconfiguration2} on setting $\eta+\tau=\vartheta$ and $\eta-\tau=\vartheta\pm \pi i$.

\section{Transmission matrices}

\p The transmission matrix that describes the scattering of a soliton with a defect will be denoted
$$T_{a\,\alpha}^{b\,\beta}(\theta),$$
where $a,b$ take the values $1$ (referring to a soliton) and $-1$ (referring to an anti-soliton), though often the soliton, anti-soliton are referred to conveniently as `+' or `-', respectively. On the other hand, $\alpha,\beta$ are integers denoting the topological charge of the defect. In all cases,
$$a+\alpha = b+\beta,$$
which represents the conservation of topological charge. In turn, this implies $\alpha$ and $\beta$ are either both odd or both even. The rapidity of the incoming soliton is $\theta$, and the transmission matrix will also depend on parameters associated with the defect.

\p
The bulk S-matrix \cite{ZZ} is given by a square $4\times 4$ matrix depending on the parameter $\Theta\equiv\Theta_{12}=(\theta_1-\theta_2)$ whose elements different from zero are:
\begin{eqnarray}\label{Smatrix}
S^{+\,+}_{+\,+}(\Theta)&=&S^{-\, -}_{-\,-}(\Theta)=\left(\frac{qx_1}{x_2}-\frac{x_2}{qx_1}\right)\rho_s(\Theta)\equiv a\, \rho_s(\Theta),\nonumber \\
S^{-\,+}_{+-}(\Theta)&=&S^{+\,-}_{-\,+}(\Theta)
=\left(\frac{x_1}{x_2}-\frac{x_2}{x_1}\right)\rho_s(\Theta)\equiv b\,\rho_s(\Theta),\nonumber \\ S^{+\,-}_{+\,-}(\Theta)&=&S^{-\,+}_{-\,+}(\Theta)
=\left(q-\frac{1}{q}\right)\rho_s(\Theta)\equiv c\,\rho_s(\Theta),
\end{eqnarray}
with
$$x_p=e^{\gamma\theta_p},\quad q=-e^{-i\pi\gamma}=e^{-4i\pi^2/\beta^2},\quad \gamma=\frac{4\pi}{\beta^2}-1.$$
The multiplicative factor $\rho_s(\Theta)$ is given by:
\begin{equation}\label{rhos}
\rho_S(\Theta)=\frac{\Gamma(1-z-\gamma)\Gamma(1+z)}{2\pi i}\,\prod^{\infty}_{k=1}R_k(\Theta)R_k(i\pi-\Theta),\quad z=\frac{i\gamma\Theta}{\pi},
\end{equation}
with
$$R_k(\Theta)=\frac{\Gamma(z+2k\gamma)\Gamma(1+z+2k\gamma)}
{\Gamma(z+(2k+1)\gamma)\Gamma(1+z+(2k-1)\gamma)}.$$
The S-matrix satisfies the usual QYBE, which, in terms of the S-matrix elements, reads
$$S_{a\,b}^{e\,d}(\Theta_{12})S_{d\,c}^{f\,i}(\Theta_{13})S_{e\,f}^{g\,h}(\Theta_{23})=
S_{b\,c}^{e\,f}(\Theta_{23})S_{a\,e}^{g\,d}(\Theta_{13})S_{d\,f}^{h\,i}(\Theta_{12}).$$
This expression should clarify the use of the index notation \eqref{Smatrix} adopted for the S-matrix elements.

\p Since the defect is purely transmitting, the T-matrix is assumed to be a solution of the following set of equations \cite{Delf94}
\begin{equation}\label{STT}
S_{a\,b}^{mn}(\Theta)\,T{_{n\alpha}^{t\beta}}(\theta_1)\,
T{_{m\beta}^{s\gamma}}(\theta_2)=T{_{b\,\alpha}^{n\beta}}(\theta_2)\,
T{_{a\,\beta}^{m\gamma}}(\theta_1)\,S_{mn}^{st}(\Theta),
\end{equation}
which expresses the compatibility between the S-matrix and the T-matrix and relies on  heuristic arguments based on factorisability and bulk integrability.
Note that while the S-matrix acts on $V\otimes V$, where $V$ is a two-dimensional space, the T-matrix acts on a $V\otimes {\cal V}$ where ${\cal V}$ is an infinite dimensional space. Also,  Latin and Greek indices refer to the finite and infinite dimensional spaces, respectively. In \cite{bcz2005}, a T-matrix was found that describes the scattering between a soliton and a type I defect. This example of a  T-matrix (for a defect in its stable configuration) is a unitary, infinite dimensional, solution of eq\eqref{STT}, conveniently summarised by
\begin{equation}\label{TImatrix}
T_{I}{^{\phantom{I}b \,\beta}_{\ a \alpha}}(\theta,\eta)=\rho_I(\theta,\eta)\left(%
 \begin{array}{cc}
   \nu^{-1/2} Q^{\alpha}\,\delta_{\alpha}^{\beta} &
    e^{\gamma(\theta-\Lambda)}\,\delta_{\alpha}^{\beta-2} \\
    e^{\gamma(\theta-\Lambda)}\,\delta_{\alpha}^{\beta+2} &
   \nu^{1/2} Q^{-\alpha}\,\delta_{\alpha}^{\beta}\\
  \end{array}
\right),
\end{equation}
with
$$e^{-\gamma\Lambda}=(-q)^{1/2}\,e^{-\gamma \eta},\quad Q=1/\sqrt{q}=e^{2i\pi^2 /\beta^2},$$
where $\eta$ is an essential constant parameter carried by the defect and $\nu$ is an inessential constant phase. The function $\rho_I(\theta,\eta)$ is not determined uniquely but a `minimal' choice was provided by Konik and LeClair \cite{Konik97}. It is:
\begin{equation}\label{}
\rho_I(\theta,\eta)=\frac{f(z)}{\sqrt{2\pi}}\,e^{-\gamma(\theta-\Lambda)/2},
\end{equation}
with
$$
f(z)=\Gamma(1/2-z)\,
\prod^{\infty}_{k=1}\frac{\Gamma(1/2+z+(2k-1)\gamma)\Gamma(1/2-z+2k\gamma)}
{\Gamma(1/2+z+2k\gamma)\Gamma(1/2-z+(2k-1)\gamma)},\quad z=\frac{i\gamma(\theta-\eta)}{\pi}.$$

\p Some of the properties of \eqref{TImatrix} have been explored in \cite{bcz2005}.

\subsection{Generating new solutions}

From a purely algebraic point of view it would be desirable to classify all solutions of eqs\eqref{STT} but for now it will be enough to demonstrate  how additional solutions might be generated.

\subsubsection{Almost tensor products}
\p
New solutions to eqs\eqref{STT} in which each defect label is replaced by a pair of labels can be manufactured by taking products of solutions of the above type. For example, taking the product of two  solutions in the following sense,
\begin{equation}\label{Tproduct}
 T_{I-I}\,{^{\,b \,\beta\,\delta}_{\,a\, \alpha\,\gamma}}(\theta,\eta_1,\eta_2)=T{_{I}}\,^{\,c \,\beta}_{\,a \alpha}(\theta,\eta_1)\,T{_{I}}\,^{\,b \,\delta}_{\,c \gamma}(\theta,\eta_2),
\end{equation}
provides another solution with two soliton labels and four defect labels.  That this satisfies eqs\eqref{STT} is easily checked diagrammatically and the combined transmission matrix represents two type I defects placed somewhere along the line, not necessarily at the same location. It is not quite a tensor product because of the matrix multiplication on the soliton labels.
 After redefining the Greek indices (by taking the new indices to be $\gamma\pm\alpha$ and $\delta\pm\beta$ then relabelling), and applying a suitable unitary transformation (to simplify the explicit dependence of the T-matrix elements on the Greek indices), the  solution \eqref{Tproduct} is equivalent to
\begin{equation}\label{TI-Imatrix}
T_{I-I}\,{^{\,b \,\beta\,\delta}_{\,a\, \alpha\,\gamma}}(\theta,\eta_1,\eta_2)=\rho_{I-I}(\theta,\eta_1,\eta_2)\,
\left(%
 \begin{array}{cc}
 t_+^+\,{^{\,\beta\,\delta}_{\, \alpha\,\gamma}}(\theta,\eta_1,\eta_2) & t_+^-\,{^{ \,\beta\,\delta}_{\, \alpha\,\gamma}}(\theta,\eta_1,\eta_2)\\
 t_-^+\,{^{ \,\beta\,\delta}_{\, \alpha\,\gamma}}(\theta,\eta_1,\eta_2) & t_-^-\,{^{ \,\beta\,\delta}_{\, \alpha\,\gamma}}(\theta,\eta_1,\eta_2)
  \end{array}
\right),
\end{equation}
with
\begin{eqnarray}\label{Tproductelements}
t_+^+\,{^{ \,\beta\,\delta}_{\, \alpha\,\gamma}}(\theta,\eta_1,\eta_2)&=&(Q^{\alpha}\,\delta_\gamma^\delta+
Q^{-\alpha}\,e^{\gamma(2\theta-\Lambda_1-\Lambda_2)} \,\delta_\gamma^{\delta+4})\,\delta_{\alpha}^{\beta},\nonumber \\
t_-^-\,{^{ \,\beta\,\delta}_{\, \alpha\,\gamma}}(\theta,\eta_1,\eta_2)&=&(Q^{-\alpha}\,\delta_\gamma^\delta+
Q^{\alpha}\,e^{\gamma(2\theta-\Lambda_1-\Lambda_2)}\,\delta_\gamma^{\delta-4})
\,\delta_{\alpha}^{\beta},\nonumber \\
t_+^-\,{^{\,\beta\,\delta}_{\, \alpha\,\gamma}}(\theta,\eta_1,\eta_2)&=&(Q^{\alpha+1}\,e^{\gamma(\theta-\Lambda_2)}\,\delta_\gamma^{\delta-2}+
Q^{-\alpha-1}\,e^{\gamma(\theta-\Lambda_1)}\,\delta_\gamma^{\delta+2})
\,\delta_{\alpha}^{\beta-2}, \nonumber \\
t_-^+\,{^{ \,\beta\,\delta}_{\, \alpha\,\gamma}}(\theta,\eta_1,\eta_2)&=&(Q^{\alpha-1}\,e^{\gamma(\theta-\Lambda_1)}\,\delta_\gamma^{\delta-2}+
Q^{-\alpha+1}\,e^{\gamma(\theta-\Lambda_2)}\,\delta_\gamma^{\delta+2})
\,\delta_{\alpha}^{\beta+2}, \nonumber
\end{eqnarray}
and
\begin{equation}\label{rhoI-I}
\rho_{I-I}(\theta,\eta_1,\eta_2)=\frac{f(z_1)
\,f(z_2)}{2\pi}\,e^{-\gamma(2\theta-\Lambda_1-\Lambda_2)/2}.
\end{equation}
In the above expression, $\alpha$ represents the total topological charge on the two defects initially and $\beta$ represents the total topological charge finally. The other Greek labels represent the exchange of charge between defects as the soliton passes between them. Clearly, multiple products of any length can be represented in a similar fashion. However, the transmission matrix representing a type-II defect is not expected to be of this type since there, as explained in \cite{cz2009}, two defects have been fused. This means  there is no soliton propagating between the two constituents to allow for an exchange of topological charge. In other words, it is expected that the transmission matrix should have just two Greek labels representing the initial and final charge on the defect. On the other hand, as was also pointed out in \cite{cz2009}, the diagonal entries of the transmission matrix are expected to dominate at both low and high rapidity (in fact, just as they do in the expressions \eqref{Tproductelements}). At first sight this appeared to be a puzzle, which was resolved only via a closer inspection of the triangle equations themselves.

\subsubsection{A new solution}
\p
One of the main results of this paper is to report a new solution to eqs\eqref{STT}, which is not of the above type, though it is similar if the `exchange' Greek labels are simply ignored (or equivalently, treated as being the same modulo two). Thus, the new solution has the form:
\begin{equation}\label{Tmatrix}
T^{b \,\beta}_{a\, \alpha}(\theta)=\rho(\theta)\,\left(%
\begin{array}{ccc}
  (a_+Q^{\alpha}+a_-Q^{-\alpha}\, x^2)\,\delta^{\beta}_{\alpha} &
  x\,(b_+Q^{\alpha}+b_-Q^{-\alpha})\,\delta^{\beta-2}_{\alpha} \\
  x\,(c_+ Q^{\alpha} + c_-Q^{-\alpha} )\,\delta^{\beta+2}_{\alpha} & ( d_+Q^{\alpha}\,x^2+d_-Q^{-\alpha} )\,\delta^{\beta}_{\alpha} \\
\end{array}%
\right),\quad x=e^{\gamma\theta},
\end{equation}
with the two constraints
$$a_\pm\, d_\pm - b_\pm \,c_\pm=0,$$
where $a_\pm$, $b_\pm$, $c_\pm$, $d_\pm$ are otherwise free (complex) constants.  Finally, $\rho(\theta)$ is a yet to be determined overall function of the soliton rapidity and other parameters.\footnote{\noindent Because of the different  dependence on the rapidity variable, $x$, in the diagonal entries of \eqref{Tmatrix} (relative to the diagonal entries of \eqref{TImatrix}), this solution was missed in \cite{bcz2005} but was not in any case relevant to the discussion given there.}

\p
The next step is to make use of crossing and unitarity in order to constrain the free constants and the form of the overall factor. The crossing relation reads
\begin{equation}\nonumber
T^{b\,\beta}_{a\,\alpha}(\theta)=\tilde{T}^{\bar a\,\beta}_{\bar b\,\alpha}(i\pi-\theta),
\end{equation}
where the matrix $\tilde{T}$ is the scattering matrix for solitons moving from the right to the left and
$$T^{b\,\beta}_{a\,\alpha}(\theta)\tilde{T}^{c\,\gamma}_{b\,\beta}(-\theta)
=\delta^{c}_{a}\delta^{\gamma}_{\alpha}.$$

\p It is useful to note that the inverse of an infinite dimensional matrix of the form
$$\rho\,\left(\begin{array}{ccc} a_\alpha\delta_\alpha^\beta & b_\alpha\delta_\alpha^{\beta -2}\\
                c_\alpha\delta_\alpha^{\beta+2}&d_\alpha \delta_\alpha^\beta\\
        \end{array}\right)
$$
is a matrix of a similar shape given by

$$\frac{1}{\rho}\,\left(\begin{array}{ccc} (d_{\alpha+2}/\Delta_\alpha)\delta_\alpha^\beta & -(b_\alpha/\Delta_\alpha)\delta_\alpha^{\beta -2}\\
                -(c_\alpha/\Delta_{\alpha-2})\delta_\alpha^{\beta+2}&(a_{\alpha-2}/\Delta_{\alpha-2}) \delta_\alpha^\beta\\
        \end{array}\right),\quad \Delta_\alpha=a_\alpha\,d_{\alpha+2}-b_\alpha \, c_{\alpha+2}.
$$
Given the specific form of the transmission matrix  \eqref{Tmatrix}, $\Delta_\alpha$ is independent of $\alpha$:
\begin{eqnarray}
\nonumber \Delta_\alpha\equiv \Delta(\theta)&=&\left(a_+d_-Q^{-2}-(b_+c_-Q^{-2}+b_-c_+Q^2)x^2+a_-d_+Q^2x^4\right)\\ &=&a_+d_-Q^{-2}\left(1-\frac{b_+c_-}{a_+d_-}\,x^2\right)\left(1-\frac{b_-c_+}{a_+d_-}\,Q^4x^2\right).
\end{eqnarray}
Crossing does not provide any further constraints on the free constants $a_\pm$, $b_\pm$, $c_\pm$,   $d_\pm$, but forces the  function $\rho(\theta)$ to satisfy the following relation:
\begin{equation}\label{rhoequation}
\rho(\theta)\rho(\theta+i\pi)\, Q^2\Delta(\theta)=1.
\end{equation}
For comparison, performing the same set of steps with the type-I T-matrix reveals
$$\Delta_I(\theta)=Q^{-2}\left(1-e^{-2\gamma\Lambda}\, Q^2 x^2\right),$$
prompting the definitions,
\begin{equation}\label{eta1eta2notation}
\frac{b_-c_+}{a_+d_-}=e^{-2\gamma{\Lambda}_1}\, Q^{-2}=-Q^{-4}\,e^{-2\gamma\eta_1},\quad
\frac{b_+c_-}{a_+d_-}=e^{-2\gamma{\Lambda}_2}\,Q^2=-e^{-2\gamma\eta_2}.
\end{equation}
Then, a solution to the crossing constraints is 
\begin{equation}\label{rhoI-IandII}
\rho(\theta)\equiv\frac{1}{\sqrt{a_+d_-}}\,\rho_{I-I}(\theta, \eta_1,\eta_2).
\end{equation}
With this choice, and up to a constant factor, the form of $\rho(\theta)$ coincides with the overall function of the T-matrix describing the scattering of a soliton by two separated type I defects with classical parameters $\eta_1,\ \eta_2$. It is important to note the following: in the case of the two type I defects the constants $\eta_1,\ \eta_2$ are real and their connections with the classical Lagrangian parameters is reasonably understood \cite{bcz2005, Bajnok}; in the present case,  $\eta_1$ and $ \eta_2$ are  complex constants whose relationship with the classical Lagrangian is not yet established.

\p Finally, since the sine-Gordon model is a unitary field theory, it is necessary to explore the consequences of unitarity for the T-matrix:
\begin{equation}\nonumber
\sum_b T^{b\,\beta}_{a\,\alpha}(\theta)\bar{T}^{b\,\beta}_{c\,\gamma}(\theta)
=\delta_{ac}\delta^{\gamma}_{\alpha}.
\end{equation}
This leads to a number of additional constraints on the parameters, which can, after making the choice $a_+=1$, and using a unitary similarity transformation to align phases in the off-diagonal terms of $T$, be expressed as follows:
\begin{equation}\label{unitaryconstraints}
a_+=d_-=1,\ a_-=b_- c_-,\ d_+=b_+c_+,\ c_-=-\bar b_+,\ c_+=-\bar b_-\, Q^{-4}, \quad \rho(\theta+i\pi)=\bar{\rho}(\theta).
\end{equation}
Then, by making use of \eqref{unitaryconstraints}, the relations \eqref{eta1eta2notation} become
\begin{equation}
 b_-\, \bar{b}_-=e^{-2\gamma{\eta}_1}, \quad b_+\, \bar{b}_+=e^{-2\gamma{\eta}_2}.\nonumber
\end{equation}
Hence, for the unitary T-matrix, the parameters $\eta_1$ and $\eta_2$ are real. This fact implies the overall function \eqref{rhoI-IandII} also satisfies the unitary constraint \eqref{unitaryconstraints}.

\p In summary, the conjectured crossing symmetric and unitary transmission matrix for the type II defect within the sine-Gordon model is
 \begin{equation}\label{TIImatrix}
T_{II}\,^{b\, \beta}_{a \,\alpha}(\theta,b_+,b_-)=\rho_{II}
\left(
\begin{array}{ccc}
  (Q^{\alpha}- b_-\, \bar{b}_+\, Q^{-\alpha}\, x^2)\,\delta^{\beta}_{\alpha} & x\,(b_+ Q^{\alpha} + b_- Q^{-\alpha})\,\delta^{\beta-2}_{\alpha} \\
  -x\,(\bar b_-\, Q^{\alpha-4} +\bar b_+\, Q^{-\alpha})  \,\delta^{\beta+2}_{\alpha} & (-b_+\bar b_-\,Q^{\alpha-4}\, x^2+Q^{-\alpha} )\,\delta^{\beta}_{\alpha} \\
\end{array}
\right),
\end{equation}
where
\begin{equation}\label{rhoII}
\rho_{II}(\theta,\eta_1,\eta_2)
=\frac{f_{II}({z}_1,{z}_2)}{2\pi}\,e^{-\gamma(\theta-\eta_1)/2}
\,e^{-\gamma(\theta-\eta_2)/2}\,e^{i\pi\gamma/2},\quad {z}_p=\frac{i\gamma(\theta-\eta_p)}{\pi},\quad p=1,2,
\end{equation}
with
\begin{eqnarray}
&&f_{II}({z}_1,{z}_2)=\Gamma(1/2-{z}_1)
\Gamma(1/2-{z}_2)\,\times\ \nonumber \\
&&\prod^{\infty}_{k=1}\frac{\Gamma(1/2+{z}_1+(2k-1)\gamma)\Gamma(1/2-{z}_1+2k\gamma)
\Gamma(1/2+{z}_2+(2k-1)\gamma)\Gamma(1/2-{z}_2+2k\gamma)}
{\Gamma(1/2+{z}_1+2k\gamma)\Gamma(1/2-{z}_1+(2k-1)\gamma)
\Gamma(1/2+{z}_2+2k\gamma)\Gamma(1/2-{z}_2+(2k-1)\gamma)}.\nonumber
\end{eqnarray}
The only significant degrees of freedom remaining are the two real constants $\eta_1$ and $\eta_2$ and the relative phase between $b_+$ and $b_-$. Notice the presence of two simple poles at $z_1=1/2$ and $z_2=1/2$ or, in terms of rapidity at
\begin{equation}\label{poles}
\theta=\eta_1+\frac{i \pi}{2\gamma},\quad \theta=\eta_2+\frac{i \pi}{2\gamma}.
\end{equation}
Following the interpretation adopted in \cite{bcz2005}, these correspond to defect `resonance' states, representing the absorption and emission of a soliton. In fact, it is straightforward to verify in the classical limit $\beta\rightarrow 0$ ($1/\gamma\rightarrow 0$) that their energies coincide with the classical energies of a soliton with rapidity $\eta_1$
or $\eta_2$, respectively. In addition, in the same limit the poles \eqref{poles} coincide with
the rapidity at which the classical soliton delay \eqref{4piconfiguration1} diverges, provided $\eta_1=\eta+\tau$ and $\eta_2=\eta-\tau$. As it happens, the T-matrix \eqref{TIImatrix} for the type II defect is supposed to describe the quantum analogue of the situation represented classically in \eqref{4piconfiguration1}, which illustrates the stable defect configuration with even topological charge labels.

\p
At this stage, it is natural to compare the T-matrix \eqref{TIImatrix}, which is supposed to describe the scattering of a soliton with a type II defect, with the T-matrix \eqref{TI-Imatrix}, which describes the scattering of a soliton with two non overlapping type I defects. In fact, the essential difference lies only in the presence of two extra defect labels for the solution \eqref{TI-Imatrix} and a choice of phase. To see this  set
$$b_-=Q^{-3}\,e^{-\gamma\eta_1}, \quad b_+=Q^{-1}\,e^{-\gamma\eta_2} $$
to find the matrix elements in \eqref{TI-Imatrix} (apart from the additional Kronecker deltas).

\p
As a small check, it is worth calculating the transmission factor for the lightest breather. It is expected to coincide with \eqref{TFlinearizedproblem} for both matrices \eqref{TI-Imatrix} and \eqref{TIImatrix}, since \eqref{TFlinearizedproblem} represents the classical counterpart in either of the cases: two type I defects or a single type II defect. This calculation does not depend on any assumptions concerning the unitarity of the T-matrix and uses only the parameter specifications contained in \eqref{eta1eta2notation}.
A breather is a bound state of a soliton-anti-soliton pair and its transmission factor is defined by the bootstrap relation
\begin{equation}\label{breatherbootstrap}
c^n_{a\,\bar{a}}\, {^n T}(\theta)\delta^{\beta}_{\alpha}=\sum_{b}T^{\bar{b}\,\gamma}_{\bar{a}\,\alpha}(\theta_{\bar{a}})
T^{b\,\beta}_{a\,\gamma}(\theta_a)c^n_{b\,\bar{b}},
\end{equation}
where
$$c^n_{+ \,-}=(-)^n c^n_{-\, +},\quad \theta_{a}=\theta+i\left(\frac{\pi}{2}-\frac{n\pi}{2\gamma}\right),\quad \theta_{\bar{a}}=\theta-i\left(\frac{\pi}{2}-\frac{n\pi}{2\gamma}\right),\quad n=1,2,\dots <{\gamma}.$$
The lightest breather corresponds to the choice $n=1$ and using \eqref{breatherbootstrap}
its transmission factor is:
\begin{equation}\label{transmissionfactorlb}
{^1 T}(\theta)=-\frac{\sinh\left(\frac{\theta-\eta_1}{2}-\frac{\pi i}{4}\right)}{\sinh\left(\frac{\theta-\eta_1}{2}+\frac{\pi i}{4}\right)}
\frac{\sinh\left(\frac{\theta-\eta_2}{2}-\frac{\pi i}{4}\right)}{\sinh\left(\frac{\theta-\eta_2}{2}+\frac{\pi i}{4}\right)}.
\end{equation}
As expected, \eqref{transmissionfactorlb} coincides with \eqref{TFlinearizedproblem} with $\eta_1=\eta+\tau,\ \eta_2=\eta-\tau$. 

\section{Reducing the T-matrix}

\subsection{The soliton-soliton S-matrix}

\p In \cite{bcz2005} it was noticed that the classical type-I defect for the sine-Gordon model behaves, in a sense explained there, as though it was `half' a soliton. In fact, the energy and momentum naturally associated with the defect is half the energy and momentum of a single soliton, and the delay experienced by a soliton travelling through the defect is half the delay experienced by a soliton overtaking another soliton with rapidity equal to the defect parameter $\eta$  ($\sigma=e^{-\eta}$). On the other hand, as shown in section (\ref{classicalsetting}), the type II defect `behaves' like a soliton for similar reasons. Hence, it is natural to wonder whether it is possible to extend such an `identification' to the quantum context, and, if so, how far it is possible to go. In this section this idea will be explored, and it will be shown how the familiar S-matrix for the sine-Gordon soliton emerges by reduction from the infinite dimensional solution \eqref{Tmatrix} of the triangular relation \eqref{STT}.

\p Clearly, the standard soliton S-matrix is finite-dimensional and will need to be located inside the infinite-dimensional solution \eqref{Tmatrix} by being the finite (four-dimensional) part of a direct sum. In order to engineer this some matrix elements within \eqref{Tmatrix} will have to be chosen to be zero by restricting the parameters. In particular, it will be necessary to choose a solution with odd topological charge labels since solitons/antisolitons carry $\pm 1$ unit of topological charge and require that there is no amplitude for transitions between topological charges $\pm 1$ and $\pm 3$, which would be allowed generically provided the incoming soliton converted to an outgoing anti-soliton (or vice-versa). To ensure these zeros it is necessary to insist that
$$ T_+^-\,{_1^3}=T_-^+\,{_{-1}^{-3}}=0,$$
which, in turn, requires
\begin{equation}\label{TSparameterconstraints}
 b_-=-b_+\, Q^2,\ c_-=-c_+\, Q^{-2}.
\end{equation}
Notice, this choice already lies outside the range of parameters \eqref{unitaryconstraints} for which the T-matrix as a whole is unitary.  However, if the following choice for the parameters is made
\begin{equation}\label{moreparameterconstraints}
a_+= d_-=1,\ a_-= d_+=-e^{-2\gamma\vartheta}Q^{-2},\ b_+=c_-=-e^{-\gamma\vartheta} Q^{-2},\ c_+=b_-=e^{-\gamma\vartheta},
 \end{equation}
then the non-zero central elements of the T-matrix become
\begin{eqnarray}
&&\qquad\qquad\qquad{T^{+}_{+}}^{+1}_{+1}(\theta)={T^{-}_{-}}^{-1}_{-1}(\theta)= a\,\hat{\rho}(\theta),\quad
\nonumber \\
&&{T^{+}_{+}}^{-1}_{-1}(\theta)= {T^{-}_{-}}^{+1}_{+1}(\theta)=b\,\hat{\rho}(\theta),\quad
{T^{-}_{+}}^{+1}_{-1}(\theta)={T^{+}_{-}}^{-1}_{+1}(\theta)= c\,\hat{\rho}(\theta),
\end{eqnarray}
where
\begin{equation}\label{rhotilde}
\hat{\rho}
(\theta)=(-Q^{-1}\,e^{\gamma(\theta-\vartheta)})\,\rho(\theta).
\end{equation}
Apart from the overall factor $\hat\rho$, these elements are precisely the non-zero elements of the
sine-Gordon S-matrix \eqref{Smatrix}. The full T-matrix may now be regarded as the direct sum of two infinite parts and a finite part and each of the parts  separately satisfies the relations \eqref{STT}. In this sense, a generally irreducible infinite dimensional solution of the triangular relation has been reduced via a special choice of parameters to a direct sum of three parts. The situation is reminiscent of one that may arise in the representation theory of $sl_2$ where, typically, a representation is infinite dimensional and irreducible. However, for certain choices of the two parameters characterizing representations a representation will split into a direct sum of a finite part together with two infinite parts.

\p
 The overall function \eqref{rhotilde} can be evaluated making use of the expression \eqref{rhoI-IandII} for the function $\rho(\theta)$  together with the choices \eqref{TSparameterconstraints} and \eqref{moreparameterconstraints} for the free constants. In fact, the expression \eqref{rhoI-IandII} has been obtained by making use only of the crossing relation, not unitarity, and therefore the constraints on the free parameters previously required by unitarity do not apply. Using the notation defined by \eqref{eta1eta2notation}  the quantities ${\eta_1}$, ${\eta_2}$ that will appear in $\hat{\rho}(\theta)$ are related to each other and given by
$$ e^{2\gamma{\eta}_1}=-Q^{-4} e^{^2\gamma\vartheta}=e^{\gamma(2\vartheta-2i\pi)-i\pi},\quad e^{2\gamma{\eta}_2}=-Q^4 e^{^2\gamma\vartheta}=e^{\gamma(2\vartheta+2i\pi)+i\pi}.$$
In other words, a suitable choice is
\begin{equation}\label{DintoSchoiceparam}
{\eta}_1=\vartheta-i\pi-\frac{i\pi}{2\gamma},\quad {\eta}_2=\vartheta+i\pi+\frac{i\pi}{2\gamma},
\end{equation}
from which it follows
$$\hat{\rho}
(\theta)=\rho_S(\theta-\vartheta)\equiv\rho_S(\Theta),$$
precisely the required identification. Notice, this has required a specific resolution of the ambiguity inherent in taking the square roots: that is, the choice $\eta_2=\bar\eta_1$, with $\vartheta$ real. It is interesting that the choices made above, though definitely inconsistent with unitarity for the T-matrix as a whole, nevertheless lead to a unitary finite piece. This works because the unitarity relation for the full T-matrix required a set of relations to hold for all labels $\alpha$ and $\beta$ whereas the labels are restricted to $\pm 1$ for the selected $4\times 4$ finite part.

\p Finally, notice that with the parameter choice \eqref{TSparameterconstraints}, and \eqref{DintoSchoiceparam}, the transmission factor for the lightest breather obtained in \eqref{transmissionfactorlb} becomes
\begin{equation}\label{transmissionfactorslb}
{^s T}(\Theta)=-\frac{\sinh\left(\frac{\Theta}{2}+\frac{i\pi}{4}\left(1+\frac{1}{\gamma}\right)\right)}{\sinh\left(\frac{\Theta}{2}-\frac{i\pi}{4}\left(1+\frac{1}{\gamma}\right)\right)}
\frac{\sinh\left(\frac{\Theta}{2}+\frac{i\pi}{4}\left(1-\frac{1}{\gamma}\right)\right)}{\sinh\left(\frac{\Theta}{2}-\frac{i\pi}{4}\left(1-\frac{1}{\gamma}\right)\right)}=\frac{\sinh(\Theta)+i\cos\frac{\pi}{2\gamma}}{\sinh(\Theta)-i\cos\frac{\pi}{2\gamma}}.
\end{equation}
This coincides with the S-matrix for a soliton and a breather whose rapidity difference is $\Theta$ \cite{ZZ}. Note, in the classical limit $\beta \rightarrow 0$ the expression \eqref{classicalsolitonbreather} is recovered.

\subsection{The breather-breather S-matrix}

As a final remark it is worth pointing out that a type II defect can also describe the lightest breather in a similar manner. Consider a transmission matrix labeled by even integers and reduce it similarly by demanding
$$ T_+^-\,{_0^2}=T_-^+\,{_{\phantom{-}0}^{-2}}=0,$$
which is achieved by setting
$$b_-=-b_+, \quad c_-=-c_+.$$
Then, setting
$a_+=d_-=1$, as before, together with
\begin{equation}\label{DintoBSchoiceparam}
 {\eta}_1=\vartheta-\frac{i\pi}{2}+\frac{i\pi}{\gamma},\quad {\eta}_2=\vartheta+\frac{i\pi}{2}-\frac{i\pi}{\gamma},
\end{equation}
a choice that is consistent with $b_+ c_-=b_-c_+$,
the breather transmission factor \eqref{transmissionfactorlb} becomes
\begin{equation}
 ^1 T(\theta)=-\frac{\sinh\left(\frac{\Theta}{2}-\frac{i\pi}{2\gamma}\right)}{\sinh\left(\frac{\Theta}{2}+\frac{i\pi}{2\gamma}\right)}
\frac{\sinh\left(\frac{\Theta}{2}+\frac{i\pi}{2}\left(1+\frac{1}{\gamma}\right)\right)}{\sinh\left(\frac{\Theta}{2}-\frac{i\pi}{2}\left(1+\frac{1}{\gamma}\right)\right)}=\frac{\sinh(\Theta)-i\sin\frac{\pi}{\gamma}}{\sinh(\Theta)+i\sin\frac{\pi}{\gamma}}.
\end{equation}
The latter is the scalar scattering matrix for a pair of lightest breathers with rapidity difference $\Theta=(\theta-\vartheta)$.

\p In this case, the non-unitary infinite dimensional pieces of the T-matrix become separated from the two equal diagonal parts $$ T_+^+\,{_0^0}=T_-^-\,{_0^0},$$
which, if in addition $a_-=d_+=-e^{-2\gamma\vartheta}Q^{-2}$, as in \eqref{moreparameterconstraints}, now represents the unitary scattering of a soliton with the lightest breather. This can be checked straightforwardly and leads to \eqref{transmissionfactorslb}, though with $\Theta \rightarrow -\Theta$ because the roles of the breather and soliton have been interchanged in the two calculations.

\p If instead of the choice \eqref{DintoBSchoiceparam} it was decided to identify
\begin{equation}
 {\eta}_1=\vartheta+\frac{i\pi}{2},\quad {\eta}_2=\vartheta-\frac{i\pi}{2},
\end{equation}
and maintain $a_-=d_+=-e^{-2\gamma\vartheta}Q^{-2}$, then
$$ ^1T(\theta)=T_+^+\,{_0^0}=T_-^-\,{_0^0}=1.$$

\p Because the breather transmission matrix \eqref{transmissionfactorlb} contains just two factors, there is no choice of defect parameters for which the defect contains the other breather states. This is because the scattering matrices for these are constructed from a product with more than two factors \cite{ZZ}. It is tempting to speculate that other breathers might be found by fusing more than two defects  in a similar manner.

\p To conclude this section, it appears that by choosing parameters carefully the type II defect includes either a pair of `half-solitons' assembled as a soliton or a pair of `half-solitons' assembled as a breather.

\section{Conclusions and outlook}

\p In this article, a type II defect within the sine-Gordon model has been investigated in a quantum context and a transmission matrix describing the interaction between a soliton and a defect has been proposed. It has been pointed out how this proposal is consistent with the classical scattering picture. Moreover, within the sine-Gordon context, the interpretation of a type II defect as two fused type I defects, a feature already mentioned in \cite{cz2009}, has been further clarified. An important yet subtle point that is worth emphasis concerns the distinction between two type I defects located at two different points on the x-axis and two type I defects compressed to the same point on the x-axis: both their classical Lagrangian descriptions and their proposed quantum transmission matrices are different. The claim is that when two defects are compressed and fused together there is no longer an option to reverse the process to restore the two separated defects. Classically, it is sufficient to notice that for separated defects, the `sewing' conditions at the location of the defect are represented by four equations, whilst for the compressed type II defect there are only three. Additionally, the quantum transmission matrix for two separated defects has two sets of defect labels, which take into account the topological charges stored at the two defects. On the other hand, the proposed transmission matrix for two fused defects has only one set of labels to keep track of the total topological charge associated with the single type II defect. In this case, the soliton that scatters with the defect cannot propagate between the two original type I defects and the only relevant quantity is the total topological charge exchanged between the incoming soliton and the fused defect. In other words, it can be argued that there is a version of `confinement' taking place in which two type I defects are permanently trapped as a type II defect.

\p Another interesting point emerging from the analysis presented here is the fact that the sine-Gordon S-matrix is embedded within a transmission matrix for suitable choices of the defect parameters. In fact, for the special choices of defect parameters the infinite-dimensional transmission matrix splits into a direct sum of three pieces, one of which is a four-dimensional matrix coinciding with the bulk soliton S-matrix. Of course, this noteworthy result is not completely unexpected. In fact, even from a classical point of view, it has been noticed that the type I defect and the type II defect resemble `half-solitons' and `solitons', respectively. These similarities  refer to the delays experienced by a soliton traversing a defect, the topological charge of the defect and the energy-momentum associated with it. The appearance of the S-matrix in the quantum case for the type II defect further supports the interpretation of this defect as a soliton-like object that can not be dismantled into its two constituent pieces.  Additionally, within the quantum context, the S-matrix for the scattering between a soliton and the lightest breather can also be recovered. This suggests, at least at the quantum level, that the defect might be seen, again with special choices for the parameters, to represent  the lightest breather. On the other hand, the situation must be different for the heavier breathers since the S-matrices that describe their scattering with a soliton do not appear in the same manner, though they can be generated using the bootstrap. Conceivably, these other S-matrices might appear embedded within infinite dimensional solutions of the triangular equations, not considered here,  in effect associated with two or more type II defects fused together. In fact, the possibility of fusing more than two type I defects deserves an investigation of its own, within both the classical and the quantum frameworks.

\p The purpose in this paper has been to investigate properties of  the type II defect within the sine-Gordon context where a link with the type I defect can be established. On the other hand, such a link does not exist for the Tzitz\'eica model. In that model a type I defect is not allowed but a type II defect is perfectly possible, as was demonstrated  in \cite{cz2009}, at least classically. Nothing is established yet in the quantum case, but, given the connections with the bulk S-matrix found for the sine-Gordon case, it would be very interesting to explore the corresponding quantum scenario for the Tzitz\'eica model. The bulk S-matrix is known \cite{IK81} and it remains to solve the appropriate triangular equations to determine candidates for a suitable transmission matrix. Actually, because of the discrepancy in the number of topological charges associated with a soliton  in the classical and quantum cases, the outcome of the analysis may be more interesting than it is in the sine-Gordon model. To recall, the permitted numbers are two classically \cite{Mik81}, and three  in the quantum field theory \cite{IK81}.

\p Although both type I and II defects have been mainly discussed within massive field theories, they can be also supported within conforml field theories, such as the Liouville model and other conformal Toda field theories. In all casses the defects are topological, in the sense they are purely transmitting, and the holomorphic and anti-holomorphic parts of the energy-momentum tensors for the fields on the right and on the left of the defect are continuous across the defect. It would be profitable to explore further these types of defect in the conformal context and to elucidate their relationship with their massive counterparts, for instance by finding the relevant defect conformal perturbation. Recently, Runkel \cite{runkel10} has tackled this problem by identifying a set of non-local charges within perturbed minimal conformal field theories containing a perturbed conformal topological defect.  However,  multi-component Toda models are not minimal and, at least some of them can support both type I and II defects. Results in this context will be presented elsewhere.

\bigskip\p{\bf Acknowledgements}

\bigskip

\p We are grateful to Peter Bowcock for discussions and to the UK Engineering and Physical Sciences Research Council for financial support under grant reference EP/F026498/1.

\appendix
\section{Solving the triangle relations}

The purpose of this appendix is to give some information concerning solutions to \eqref{STT}. Some of the details are similar to the arguments made in \cite{bcz2005} but, since there are critical differences, necessary details are repeated here. A typical transmission matrix is represented, using a block structure for the soliton labels, and taking into account the preservation of topological charge, as follows:
\begin{equation}
    T_\alpha^\beta=\left(
       \begin{array}{cc}
         A_\alpha^\beta & B_\alpha^\beta \\
         C_\alpha^\beta & D_\alpha^\beta \\
       \end{array}
     \right),
\end{equation}
with
$$A^\beta_\alpha=a_\alpha \delta^\beta_\alpha,\quad D^\beta_\alpha=a_\alpha \delta^\beta_\alpha,\quad B^\beta_\alpha=b_\alpha \delta^{\beta-2}_\alpha,\quad C^\beta_\alpha=c_\alpha \delta^{\beta+2}_\alpha.$$
Then, the relations \eqref{STT} split into several groups of equations, which will be tackled in turn. The first group includes the following relations
\begin{equation}\label{group1}
A_1A_2=A_2A_1,\quad D_1D_2=D_2D_1,\quad B_1B_2=B_2B_1,\quad C_1C_2=C_2C_1,
\end{equation}
where $\alpha$ and $\beta$ are either both even or both odd integers (positive or negative) and the subscript refers to the rapidities $\theta_1$ and $\theta_2$.  Since $A_i$ and $D_i,\ i=1,2$ are diagonal the first two relations are always satisfied, while the other two require a factorised dependence for $B_i$ and $C_i$ on the defect label $\alpha$ and the rapidities $\theta_i$. Thus,
\begin{equation}\label{}
b_\alpha=g(\alpha)b_0(\theta),\quad c_\alpha=\tilde{g}(\alpha)c_0(\theta),
\end{equation}
that is the $\theta$ and $\alpha$ dependence factorise.

\p
The next group is
\begin{equation}\label{group2}
b(A_1D_2-D_2A_1)=c(C_2 B_1-C_1 B_2),\quad b(A_2D_1-D_1A_2)=c(B_1 C_2-B_2 C_1),
\end{equation}
where $b$ and $c$ are $S$-matrix elements from \eqref{Smatrix}. These imply
$$c_0(\theta)=b_0(\theta)\equiv b_0(\theta).$$
Then, the following group
\begin{equation}\label{group3}
b(B_1 C_2-C_2 B_1)=c(D_2 A_1-D_1 A_2),\quad b(B_2 C_1-C_1 B_2)=c(A_1D_2-A_2D_1),
\end{equation}
reduces to the following constraint
\begin{equation}\label{group3constraint}
   b\, b_0^1b_0^2\,(g(\alpha)\tilde{g}(\alpha+2)-\tilde{g}(\alpha)g(\alpha-2))=c\, (a_\alpha^1 d_\alpha^2-a_\alpha^2d_\alpha^1),
\end{equation}
where superscripts 1,2 have been used to distinguish rapidities.

\p
Another group of equations is
\begin{eqnarray}
  &a A_1 B_2=b B_2 A_1+c A_2 B_1,\quad &a B_1 A_2=b A_2 B_1+c B_2 A_1,\label{group4a} \\
  &a A_2 C_1=b C_1 A_2+c A_1 C_2,\quad &a C_2 A_1=b A_1 C_2+c C_1 A_2,\label{group4b}.
\end{eqnarray}
However, \eqref{group4a} and \eqref{group4b} lead to the same constraints and it is only necessary to consider one of them \eqref{group4a}.
The two equations in \eqref{group4a} can be combined to eliminate $B_1$ and $B_2$ leading to a quadratic recurrence relation for the components of $A_i$:
\begin{equation}\label{group4constraint1}
(b^2-c^2)\, a^1_{\alpha+2}a^2_\alpha+a^2\, a^1_\alpha a^2_{\alpha+2}=ab\, (a^1_\alpha a^2_\alpha+a^2_{\alpha+2}a^1_{\alpha+2}).
\end{equation}
Suppose that the solution for $a_\alpha$ factorizes,
\begin{equation}\label{achoicefac}
a_\alpha(\theta)=f(\alpha)\, a_0(\theta).
\end{equation}
Then, equation \eqref{group4constraint1} implies
$$f(\alpha)=q^{\pm\alpha/2},$$
and the use of one of the expressions in \eqref{group4a} also implies
\begin{equation}\label{}
 \mbox{if } f(\alpha)=q^{\alpha/2}, \quad b_0(\theta)=\frac{a_0(\theta)}{x},\qquad
 \mbox{if } f(\alpha)=q^{-\alpha/2}, \quad b_0(\theta)=x a_0(\theta).
\end{equation}
Indeed, this was the choice originally made in \cite{Konik97} and supported by the semiclassical arguments developed in \cite{bcz2005}. The evidence suggests it is the appropriate solution for a type I defect.

\p
However, the choice \eqref{achoicefac} is not inevitable.
A more interesting solution to \eqref{group4constraint1}  in the context of type II defects is
\begin{equation}\label{achoice}
a_\alpha(\theta)=(a_+\,q^{-\alpha/2}+a_- \, q^{\alpha/2}x^2)\rho(\theta)\equiv (a_+\,Q^\alpha+a_-\,Q^{-\alpha} x^2)\rho(\theta).
\end{equation}
Then, the use of one of the expressions in \eqref{group4a} implies
\begin{equation}\label{}
b_0(\theta)=x \rho(\theta).
\end{equation}
The remaining group of equations to analyze is
\begin{eqnarray}
  &a D_1 C_2=b C_2 D_1+c D_2 C_1,\quad &a C_1 D_2=b D_2 C_1+c C_2 D_1,\label{group5a} \\
  &a D_2 B_1=b B_1 D_2+c D_1 B_2,\quad &a B_2 D_1=b D_1 B_2+c B_1 D_2,\label{group5b}.
\end{eqnarray}
Here, the $D_i$ are playing the the same role as $A_i$ in the previous group and therefore there is a similar collection of solutions, the relevant one in the present context being
\begin{equation}\label{dchoice}
d_\alpha(\theta)=(d_+\,q^{-\alpha/2} x^2+d- \, q^{\alpha/2})\rho(\theta)\equiv (d_+\,Q^\alpha x^2+d_-\,Q^{-\alpha})\rho(\theta).
\end{equation}

\p Finally, the relation \eqref{group3constraint} should determined the functions $g(\alpha)$,
$\tilde{g}(\alpha)$,
 $$c(a_- d_- q^{\alpha}-a_+d_+ q^{-\alpha}) =g(\alpha)\tilde{g}(\alpha+2)-\tilde{g}(\alpha)g(\alpha-2).$$
Again, this is a quadratic recurrence relation and the relevant solutions are those supplied in the text, eq\eqref{Tmatrix}, together with the constraints on the parameters given there.

\p There are other solutions to \eqref{STT}, and some may be associated with the fusing of three or more type I defects; these have not been classified systematically.


\begin{thebibliography}{99}

\bibitem{Delf94}
G.~Delfino, G.~Mussardo, and P.~Simonetti, {\it Statistical models with
a line of defect}, {\it Phys. Lett.} B {\bf 328} 123;\hepth{9403049}.

G.~Delfino, G.~ Mussardo and  P.~Simonetti,
{\it Scattering theory and correlation functions in statistical
models with a line of defect}, {\it Nucl. Phys.}  B {\bf 432} 518; \hepth{9409076}.

\bibitem{Konik97}
R. ~Konik and A.~LeClair, {\it  Purely transmitting defect field
theories}, {\it Nucl. Phys.} B {\bf 538} 587; \hepth{9703085}.


\bibitem{bcz2003}
  P.~Bowcock, E.~Corrigan and C.~Zambon,
 {\it Classically integrable field theories with defects},
  Int.\ J.\ Mod.\ Phys.\  A {\bf 19S2} (2004) 82; arXiv:hep-th/0305022.

  P.~Bowcock, E.~Corrigan and C.~Zambon,
  {\it Affine Toda field theories with defects},
  JHEP {\bf 0401} (2004) 056; arXiv:hep-th/0401020.

\bibitem{bcz2005}
  P.~Bowcock, E.~Corrigan and C.~Zambon,
  {\it Some aspects of jump-defects in the quantum sine-Gordon model},
  JHEP {\bf 0508} (2005) 023; arXiv:hep-th/0506169.


\bibitem{others}
  V.~Caudrelier,
  {\it On a systematic approach to defects in classical integrable field theories},
  Int.\ J.\ Geom.\ Meth.\ Mod.\ Phys.\  {\bf 5} (2008) 1085.

  I.~Habibullin and A.~Kundu,
  {\it Quantum and classical integrable sine-Gordon model with defect},
  Nucl.\ Phys.\  B {\bf 795} (2008) 549; arXiv:0709.4611 [hep-th].

P. Bowcock and J. M. Umpleby,
\emph{Defects and Dressed Boundaries in Complex sine-Gordon theory},
JHEP \textbf{0901} (2009) 008; arXiv:0805.3668[hep-th].


  F.~Nemes,
  {\it Semiclassical analysis of defect sine-Gordon theory},
  arXiv:0909.3268 [hep-th].
 
\bibitem{Bajnok}
  Z.~Bajnok and Z.~Simon,
  {\it Solving topological defects via fusion},
  Nucl.\ Phys.\  B {\bf 802} (2008) 307; arXiv:0712.4292 [hep-th].
 

\bibitem{Caudrelier:2004hj}
  V.~Caudrelier, M.~Mintchev, E.~Ragoucy and P.~Sorba,
 {\it Reflection-transmission quantum Yang-Baxter equations},
  J.\ Phys.\ A  {\bf 38} (2005) 3431; arXiv:hep-th/0412159.


\bibitem{Bajnok:2009hp}
  Z.~Bajnok and O.~el Deeb,
  {\it Form factors in the presence of integrable defects},
  Nucl.\ Phys.\  B {\bf 832} (2010) 500; arXiv:0909.3200 [hep-th].


\bibitem{cz2009} E. Corrigan and C. Zambon,
  {\it A new class of integrable defects},
  J.\ Phys.\ A  {\bf 42} (2009) 475203; arXiv:0908.3126 [hep-th].




\bibitem{ZZ}
A.~B.~Zamolodchikov and Al.~B. Zamolodchikov,
{\it Factorized S-matrices in two dimensions as the exact solutions of certain relativistic quantum field theory models}, Ann. Phys. {\bf 120} (1979), 253.



 \bibitem{IK81}
A. G. Izergin and V. E. Korepin, \emph{The Inverse Scattering Method
Approach to the Quantum Shabat-Mikhailov Model}, Commun. Math. Phys.
\textbf{79} (1981) 303.

F. A. Smirnov, \emph{Exact S matrices for $\phi_{(1,2)}$ perturbed minimal
models of conformal field theory}, Int. J. Mod. Phys. \textbf{A6} (1991)
1407.

\bibitem{Mik81} A. V. Mikhailov, \emph{The reduction problem and the
inverse scattering
method}, Physica {\bf D3}, (1981) 73.

N. J. MacKay and W. A. McGhee, \emph{Affine Toda solitons and
automorphisms of Dynkin diagrams}, Int. Journ. Mod. Phys {\bf A8} (1993)
2791; erratum-ibid {\bf A8} 3830; hep-th/9208057.

\bibitem{runkel10}
  I.~Runkel,
  {\it Non-local conserved charges from defects in perturbed conformal field
  theory}, arXiv:1004.1909.

\end{thebibliography}
\end{document}